\documentclass{article}
\usepackage[preprint]{spconf}
\usepackage{amsmath,graphicx,booktabs,amssymb,url}
\usepackage[T1]{fontenc}

\ninept

\title{Self-Distilled Pronunciation and Accent Control for Neural Text-to-Speech}

\name{Shuhei Kato}
\address{Independent Researcher, Japan \quad KOWRO Co., Ltd., Japan \\ \texttt{shuhei@shuheikato.info}}

\toappear{Preprint. Submitted to ICASSP 2027. This version corrects the submitted text and adds the release.}
\begin{document}
\maketitle

\begin{abstract}
Text-to-speech that reads raw text has no lexicon: a rare word is read as guessed, and a native Japanese listener accepts a word only if its reading \emph{and} pitch accent are both right. A known remedy installs a reading-and-accent channel into a released model, but it needs many recordings. This paper removes the recordings: the frozen backbone reads a sentence containing a \emph{common} word it already says correctly, and that output serves as the teacher for the same sentence with the word replaced by an annotated reading with its pitch accent. Screened raters judged the tag right on 0.80 to 0.93 of unseen difficult words on four backbones spanning autoregressive, diffusion, and encoder–decoder synthesis; plain kana, which cannot express an accent, got 0.38 to 0.60. On words needing no edit, naturalness is non-inferior on one backbone; on the other three, listeners prefer the unedited rendition by 0.19 to 0.26.
\end{abstract}


\section{Introduction}
\label{sec:intro}

Conventional cascade text-to-speech (TTS) resolves pronunciation before synthesis: a grapheme-to-phoneme (G2P) front-end consults a lexicon and adopts pre-registered pronunciational rules. End-to-end TTS trained on raw text drops G2P in exchange for improved naturalness and expressiveness. A rare word, a proper noun, or a domain term is rendered as the model guesses, and the only recourse is to change the model.

Japanese has a further problem. What matters is not only which morae are produced but which of them carries the pitch accent. Accent is lexically specified, and getting it wrong is not merely unnatural. The listener searches the lexicon for a word that was never uttered, so the cost falls on recognition.

Input-side control is not new: Kurihara et al.~\cite{kurihara2021prosodic} put prosodic symbols into the input of neural sequence-to-sequence TTS. UtterTune~\cite{kato2025uttertune} realizes them for a modern large language model (LLM)-based backbone by wrapping the target word in two special tokens carrying a katakana (phonogram) reading with pitch accent. Sarashina2.2-TTS~\cite{sarashina2026tts} describes the same mechanism, obtained from ${\approx}4{,}000$\,h of supervised fine-tuning and absent from the released weights.

Some methods intervene on each target word. GRAFT~\cite{asonitis2026graft} fine-tunes the base model to condition the pronunciation of one word on a spoken sample of it. Training-free editing exists for released models: SonoEdit~\cite{singh2026sonoedit} and FlowEdit~\cite{singh2026flowedit} edit one word's pronunciation into the weights or the text embedding; SpeechSplice~\cite{kurihara2026speechsplice} replaces the encoder's hidden states for the word in an encoder--decoder TTS. The first three require an audio clip for each target word; SpeechSplice needs none but an encoder--decoder backbone. None of the four addresses accent.

Mixing phoneme or reading input into the training set, as in CosyVoice~3's pronunciation inpainting~\cite{du2025cosyvoice3} and GLM-TTS's hybrid phoneme--text input~\cite{cui2025glmtts}, gives a base model built-in controllability, but cannot be added to a model already released. Preference optimization on the model's own samples, ranked by an automatic speech recognition (ASR) system, improves reading accuracy at the token level~\cite{kotoge2025tkto}; it repairs the default reading of the words it was trained on and adds no channel for a user to write a reading or an accent.

Of the methods listed above, only Kurihara et al.~\cite{kurihara2021prosodic} and UtterTune~\cite{kato2025uttertune} have evaluated the controllability of pitch accent. Both were obtained by training models on recordings with accent annotations. Kurihara et al. used $7{,}596$ pairs from JSUT~\cite{sonobe2017jsut}, while UtterTune used $15{,}097$ hand-corrected pairs from JSUT/JVS~\cite{sonobe2017jsut,takamichi2019jvs}. Each was shown on a single architecture, sequence-to-sequence and LLM-based (CosyVoice~2~\cite{du2024cosyvoice2}), respectively. What is unresolved for a released model is the \emph{cost} of installing the channel afterward and its \emph{reach}.

This paper eliminates the recording cost entirely. Given a reading spelled out, the model can already say it; what it lacks is a channel that carries the reading and the accent together. We install that channel by distilling the model against itself. What is new is the training pair, not the adapter or the tag.

\noindent\textbf{Contributions.}
(i) A self-distillation procedure that installs reading and accent control into a released TTS backbone from a text set and a public front-end, with no recorded speech (\S\ref{sec:method}).
(ii) Evidence that the principle is architecture-agnostic, with its failure mode and its cost: one recipe, registered in advance and applied unchanged to four backbones, delivers reading and accent together where plain kana cannot (\S\ref{sec:exp}).
(iii) The deployment form that makes it usable: register a word once, and every later occurrence is corrected automatically, accent included (\S\ref{sec:deploy}).

\noindent\textbf{Non-claims.} We claim no novelty for low-rank adaptation or for phoneme-mode tag tokens as such~\cite{eskimez2024e2tts}, and no language independence; sentence-level character error rate is not used as a correctness metric (\S\ref{sec:exp}).

\section{Method}
\label{sec:method}

\subsection{Control signal}
\label{sec:control_signal}
The target word is replaced, in the input text, by \texttt{<PHON\_START>}\allowbreak\,\emph{katakana with pitch accent}\,\allowbreak\texttt{<PHON\_END>}. An apostrophe marks the mora after which the pitch falls (the accent nucleus); a slash marks an accent-phrase boundary; an unmarked reading is flat (\emph{heiban}). The notation follows JEITA IT-4006~\cite{jeita2010it4006}. Only the tag spelling changes across backbones.

\subsection{Self-distilled teachers}
Let $w$ be a \emph{common} word that the base pronounces accurately, and let $c(w)$ denote a natural carrier sentence that includes $w$ in standard orthography. The frozen base synthesizes $c(w)$, and we keep \emph{its own output} as the target. The student input $\tilde c(w)$ is $c(w)$ with $w$ replaced by its reading with pitch accent. Distilling
\[
\tilde c(w) \;\longmapsto\; \text{base output for } c(w)
\]
teaches the model to honor an explicit reading and accent instruction while changing nothing else. The teacher and the student describe the same utterance. No recorded speech or its annotation enters the adapter's training data; what the recipe does use is public and automatic: a G2P front-end supplies the readings, and an automatic speech recognizer (ASR) checks them. Training a TTS model on its own ASR-verified output is not new~\cite{asaria2026reliable}, and distilling a model that sees one input into the same model given another is context distillation~\cite{snell2022context}; here the two inputs are two spellings of one utterance, and what is internalized is the meaning of a notation.

\textbf{One teacher rule.} A take is retained only if it both clears the degeneration guard in \S\ref{sec:guard} \emph{and} the reference ASR system (kana-whisper, \S\ref{sec:exp}) detects the target word's reading within it; otherwise the seed is bumped, and the carrier is drawn again, up to five takes. A carrier with no passing take is dropped and logged with its transcripts. No best-of-$n$, no error-rate threshold, no per-word cap.

\textbf{One rate rule.} A retained teacher is discarded if its speech exceeds $0.22$\,s per character of the plain carrier, as computed from the backbone's native representation (speech tokens, latent or codec frames, or the waveform). The threshold is a single number on every backbone, regardless of what it removes; \S\ref{sec:guard} explains why it exists. Out of $23{,}549$ carriers, the teacher rule retains $22{,}729$ for Sarashina, $21{,}395$ for CosyVoice~2, $22{,}548$ for Irodori, and $21{,}449$ for T5Gemma; the rate rule subsequently reduces these to $12{,}870$, $13{,}751$, $13{,}901$, and $18{,}781$, respectively. Dropped carriers stay out; they are not replaced.

\textbf{One speaker rule.} Where the backbone synthesizes from text alone (Sarashina, T5Gemma), no speaker enters: the teacher is the base's own rendition of the plain carrier, and the identity is provided at synthesis time by the frozen stack from a reference the adapter was never exposed to. When that is not possible (Irodori, CosyVoice~2), the \emph{same} $2{,}000$ synthetic voices—sampled from an earlier public Irodori release without a reference and stored once—are rotated across the carriers: as reference latents for Irodori, as prompt clips with their own transcripts for CosyVoice~2. No recording enters at any point.

\subsection{The text set, and why odaka shapes it}
The word inventory contains $471$ common words, each in up to $50$ frame-built carriers ($23{,}549$ in all), balanced across ten semantic domains, mora-initial consonant rows, and all four accent types. One type dictates the design: for \emph{odaka} words, the fall occurs on the \emph{following} particle, so it is inaudible in isolation. Every odaka carrier therefore places a particle immediately after the target. Odaka, the hardest type, is deliberately over-sampled ($15\%$). No word shares a surface form or a canonicalized reading with any evaluation target. The training and evaluation vocabularies are disjoint by construction.

\subsection{Training}
What installs the channel is the training pair above: the tagged student input paired with the base's own output for the plain carrier. The parameterization of the update is incidental: any adaptation that leaves the codec and vocoder frozen would serve. Here, Sarashina and CosyVoice~2 use rank-16 LoRA~\cite{hu2022lora} ($\alpha=64$) on the query, key, value, and output projections plus \emph{only the two tag-token embedding rows}; T5Gemma uses rank-8 LoRA ($\alpha=16$) on the attention and MLP projections; Irodori is fully fine-tuned. The codec and vocoder are never touched ($20{,}000$ steps on Sarashina, $1{,}000$ and $2{,}000$ on CosyVoice~2 and Irodori, $600$ token-budget steps on T5Gemma; \S\ref{sec:exp}).

\textbf{Tag tokens.} Sarashina ships the two tag tokens; CosyVoice~2 and Irodori get two new embedding rows; T5Gemma-TTS uses two reserved tokens.

\section{Deployment: register once, applied automatically}
\label{sec:deploy}

The deployed system maintains a \emph{registry} that maps a surface form to its accented reading --- either \{\emph{reading}, \emph{accent index}\} or the accented katakana written directly --- and applies it via a router at synthesis time. The router runs the public front-end over the \emph{whole} sentence, matches each registered surface \emph{at morpheme boundaries}, and then
\begin{enumerate}\itemsep2pt
  \item wraps only the matched span with the tags, carrying that entry's reading with pitch accent;
  \item \textbf{absorbs a following suffix} into the wrapper when the front-end attaches one, so
        that the tag never abuts raw unconverted orthography;
  \item \textbf{falls back to the unedited base} when no registered surface aligns with a morpheme
        boundary, and records that the edit did not fire.
\end{enumerate}

\textbf{A word is registered once, with its accent, and every later occurrence in any sentence is corrected automatically}, with no per-sentence markup, no retraining, and, unlike per-word editors~\cite{singh2026sonoedit,singh2026flowedit}, no exemplar.

\section{Experiments}
\label{sec:exp}

\begin{figure}[t]
\centering
\includegraphics[width=\columnwidth]{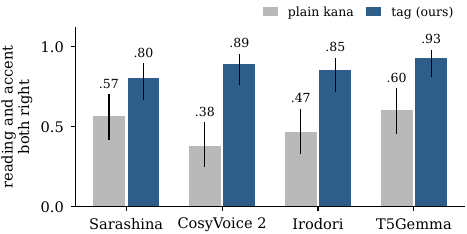}
\caption{Share of $50$ unseen difficult words that screened raters judge right in \emph{both} reading and accent, plain kana against the tag, judged blind in the same cell (paired words; whiskers: Wilson $95\%$). Kana cannot express an accent. Four raters, majority judgment.}
\label{fig:both}
\end{figure}

\subsection{Setup}
We evaluate four backbones spanning generation paradigm, codec, and tokenizer: the autoregressive LLM-based Sarashina2.2-TTS~\cite{sarashina2026tts} and CosyVoice~2~\cite{du2024cosyvoice2}, the diffusion model Irodori-TTS-500M-v3~\cite{aratako2026irodori}, and the encoder--decoder T5Gemma-TTS~\cite{arata2026t5gemmatts}. The text set, teacher rule, speaker rule, item sets, voices, guard, tests, rater rules, and exclusions were finalized in a pre-registration before any teacher audio was reviewed, and they are applied in the same way to the four. The registration also holds the full execution record, with every amendment dated, and is released with its commit history.

\textbf{Earlier rounds.} Two earlier rounds are superseded and not pooled; their full records stand in the registration. The second, registered in the same way as this one but with ten carriers per word and no rate rule, failed on Sarashina, whose adapter ran away (\S\ref{sec:guard}). The round described here modifies only the teacher set (up to $50$ carriers per word and the rate rule) across all four backbones, a change fixed before any of its outputs existed; all judgments are gathered again from scratch.

\textbf{Checkpoints.} On CosyVoice~2, Irodori, and T5Gemma, the step count is chosen on validation data that shares no items with any reported set ($105$ selection items of the objective set and $39$ sentences around common words), by a rule fixed before the adapters existed: lowest sentence-level kana CER of the tag arm, ties within $0.005$ to the higher UTMOSv2. As a result, we adopt $1{,}000$, $2{,}000$, $600$ steps on CosyVoice~2, Irodori, T5Gemma, respectively; Sarashina keeps its registered $20{,}000$ steps.

Reading is scored as canonicalized target-word hit under one pass of a kana-constrained ASR (kana-whisper, released with~\cite{sarashina2026tts}), with audio resampled to $16$\,kHz; the Kana column in Table~\ref{tab:reading} serves as a check that the ASR correctly reads the output of each backbone ($\ge 0.84$ for all four).

\textbf{Item sets.} Two subjective sets are prepared for different purposes. The $50$-word \emph{difficult} set consists of hard-to-read words, designed to evaluate reading and accent. A $50$-word \emph{easy} set comprises words the base already reads, serving the naturalness study alone (\S\ref{sec:nat}) to measure degradation. The objective set comes from the Joyo Kanji Yomi Benchmark~\cite{joyo2026} filtered by the following procedure: 1) keep every sentence on which a public dictionary G2P (pyopenjtalk~\cite{pyopenjtalk}) misreads the marked span; 2) drop any whose word or reading occurs in the training inventory; 3) replace the marked span with its katakana reading and confirm, by running the G2P on the substituted sentence, that the intended reading results. This leaves $424$ items, with no sampling and nothing about our models in the selection; a $319$-item test half is split off by connected component of the (word, reading) graph before any adapter was measured.

Each set is compared by representing the target word in different ways. The \emph{kana} condition replaces the target word with katakana, expected to correct the reading, not the pitch accent. The \emph{tag} condition substitutes the target word with the proposed control signal (\S\ref{sec:control_signal}). The objective set has the third condition: no edit (raw text).

\textbf{Reference voices.} Every result on the test half, and every listening and naturalness stimulus, is synthesized under one of $42$ held-out voices per item, assigned by hash before any audio existed: $42$ ReazonSpeech~\cite{yin2023reazonspeech} large-v2 speakers chosen as cluster centroids of a speaker-embedding space, considering acoustic conditions and the accuracy of the transcription. None of them appears in any training or teacher-generation stage.

\subsection{Rejecting degenerate output, identical in every arm}
\label{sec:guard}
One guard rejects degenerate output in every arm. Each backbone sometimes produces non-speech output (silence, a repeated mora, or a stretch that hits the token limit), which a reading metric would count as misses. One implementation is applied to every arm of every backbone: a take is rejected if it is under $0.5$\,s, reaches the generator's length cap, is silent, is too quiet to hear, or transcribes as fewer than two morae or as one mora repeated six times or more. The seed is bumped, and the \emph{first} valid take is kept, up to six. Never the best of several because that would be a take selection. The retry record is released per item.

What the guard sees is a result in itself. In the earlier round, the Sarashina tag arm needed a redraw on $60$ of $319$ objective items, with $3$ for no edit and $3$ for kana. On $20$ of them, all six takes hit the token cap. Six teacher-set variants trained in parallel (exploratory) isolated one property: the base's text-only continuations are often slow ($56\%$ of teachers above $0.22$\,s per character), and an adapter trained on them lengthens every output. Dropping those teachers removes the runaway ($1$ redraw in $100$ items, against $11$ to $18$ for variants that moved the tag off the sentence start, removed repeated-token runs, or skipped the reading check); five times the carriers without the rule do not ($41$ of $319$). The variant was picked from the first $100$ test items, so Sarashina's row in Table~\ref{tab:reading} is marked; for the other three backbones, the rule was fixed before any of their outputs were available. With the rule, the Sarashina tag arm redrew $4$ of $319$ items and exhausted the takes on none; CosyVoice~2 and Irodori redrew none; T5Gemma redrew $32$, and $7$ of its outputs stayed inaudible and are scored as misses. The rule is not finely tuned (a threshold of $0.32$\,s yields the same result).

\subsection{Does control transfer to unseen words?}
\label{sec:transfer}
Yes, on every backbone: on the objective $319$-item test half, words that appear in no training carrier, the tag reads $0.28$ to $0.50$ above no edit (Table~\ref{tab:reading}; one synthesis recipe and one ASR pass for every cell).

\begin{table}[t]
\caption{Target-reading accuracy on $319$ unseen rare words (Joyo benchmark items a public dictionary G2P misreads; \S\ref{sec:exp}), each under one of $42$ held-out voices fixed by hash. Intervals are $95\%$ bootstrap ($10{,}000$ resamples), clustered by word ($210$ word-reading components), with exact McNemar $p$-values. ``NE'' (no edit) is the raw text input; ``Kana'' writes the reading in plain kana: the phoneme-level input baseline on the same backbone, which cannot express an accent. UT, the released UtterTune adapter, exists only for CosyVoice~2 ($15{,}097$ recorded pairs); every ``Ours'' used no recorded speech; all arms share items, voices, seeds, and guard (\S\ref{sec:guard}).}
\label{tab:reading}
\centering\small\setlength{\tabcolsep}{1.4pt}
\begin{tabular}{lccccl}
\toprule
Backbone & NE & Kana & UT & Ours & Ours$-$Kana, $p$ \\
\midrule
Sarashina$^\dagger$ & .511 & .843 & --- & .875 & $+.031\,[-.016,+.080]$, .20 \\
CosyVoice~2 & .282 & .850 & .856 & .784 & $-.066\,[-.113,-.019]$, .009 \\
Irodori    & .586 & .909 & --- & .868 & $-.041\,[-.075,-.009]$, .019 \\
T5Gemma    & .436 & .906 & --- & .793 & $-.113\,[-.154,-.073]$, $10^{-7}$ \\
\bottomrule
\end{tabular}
\par\vspace{2pt}\parbox{\columnwidth}{\footnotesize $^\dagger$The rate rule was identified on the first $100$ of these items (\S\ref{sec:guard}). On the untouched $219$: Ours $.904$, Kana $.868$ ($+.037\,[-.026,+.100]$).}
\end{table}

Against no edit, the tag gains $+0.364$ $[+0.291,+0.436]$ on Sarashina, $+0.502$ $[+0.424,+0.576]$ on CosyVoice~2, $+0.282$ $[+0.215,+0.351]$ on Irodori and $+0.357$ $[+0.277,+0.436]$ on T5Gemma ($p<10^{-19}$ throughout): reading control reaches words it never saw tagged in every paradigm. Against plain kana, it reads level on Sarashina and $0.04$ to $0.11$ lower on the other three. T5Gemma's misses concentrate on a few reference voices ($32$ of $66$ on $8$ of the $42$). But this score ignores accent: it counts as right a rendition whose morae are right and whose accent is wrong, which to a native listener is simply wrong. Writing an accent into the kana does not help: to the frozen base, an apostrophe is punctuation, not a nucleus. Giving that notation a meaning is what the distillation does. The listening evaluation (\S\ref{sec:listening}), which scores reading and accent together, is where the comparison with kana is decided. The ASR score is also conservative for both arms: on the listening stimuli, where both measures exist, raters judged the reading wrong in $2$ of $200$ tagged clips and $0$ of $200$ kana clips (majority of four); the ASR missed the target in $42$ and $35$ clips, respectively.

\textbf{What zero recorded data buys and costs.} Only CosyVoice~2 has a recorded-data adapter to compare with: the released UtterTune, trained on $15{,}097$ hand-corrected recorded pairs. Ours uses no recording and no label; on the same $319$ items and voices it trails UtterTune by $0.072$ $[-0.120,-0.024]$ ($p=0.003$; Table~\ref{tab:reading}). The other three backbones have no such corpus or adapter.

SpeechSplice~\cite{kurihara2026speechsplice}, a concurrent work, edits pronunciation on T5Gemma-TTS without training by replacing encoder hidden states. Ours needs training but does not depend on such an intervention point.

\subsection{Listening evaluation}
\label{sec:listening}
On all four backbones, the tag provides both reading and accent, whereas kana does not (Table~\ref{tab:subj}, Fig.~\ref{fig:both}). A rendition is correct for a native listener only when both the reading and the accent are right. The raters select ``matches the prescribed accent'' in that case, and its share is the outcome. It is judged on all four backbones, with the same comparator: the tag versus plain kana on the $50$ difficult words, presented as whole sentences, each under the item's held-out voice ($27$ to $31$ voices per cell), and on Sarashina also versus no edit; a total of $450$ stimuli, rated blind and presented in randomized order with the specified accent displayed, plus $23$ duplicate items and $8$ known-answer catch trials, for $481$ items per rater.

Accent realization is a correctness judgment, so ability is gated by a known-answer screening test rather than by credentials: $20$ items, eligibility requiring a score of $18$ or higher, binding on the first try, all recruitment pathways using the same test, each candidate compensated based on their score, and no evaluator informed which version came from which system. Of $34$ candidates ($33$ through a crowdsourcing platform by open call; $1$ by direct approach), six passed, at $20$, $19$, $19$, $18$, $18$ and $18$.

Each admitted rater judges everything. A rater below $6/8$ on the catch trials, or answering faster than the shortest stimulus, is excluded and reported. Fleiss' $\kappa$ is calculated over all stimuli, ties included. Tied items are excluded for aggregation, and their number is reported. Five contrasts are confirmatory (McNemar exact, Holm-corrected, one family): tag against kana on each of the four backbones, and tag against no edit on Sarashina. The stimulus and screening definitions, evaluation scripts, text set, teacher sets, adapter weights (under each backbone's license), and training code are at \url{https://github.com/shuheikatoinfo/UtterTune-SelfDistill}.

\begin{table}[t]
\caption{Subjective results (four raters; $102$ naturalness sittings). Left: share of the $50$ difficult words judged right in reading \emph{and} accent; risk difference with $95\%$ CI ($10{,}000$ resamples, here and on the right); exact McNemar $p$, Holm-corrected over the five confirmatory contrasts ($^\ddagger$tag against no edit, whose share is in the Kana column). Right: naturalness on the easy set, share preferring the edited (tag) rendition minus share preferring the no-edit one (negative: the edit costs naturalness); $^\ast$lower limit within the registered margin of $-0.15$.}
\label{tab:subj}
\centering\footnotesize\setlength{\tabcolsep}{1.5pt}
\begin{tabular}{lcclcl}
\toprule
 & \multicolumn{4}{c}{Reading and accent both right} & \multicolumn{1}{c}{Naturalness} \\
\cmidrule(lr){2-5}\cmidrule(lr){6-6}
 & Kana & Tag & \multicolumn{1}{c}{Tag$-$Kana} & $p$ & \multicolumn{1}{c}{Tag$-$No edit} \\
\midrule
Sarashina   & .57 & .80 & $+.24\,[+.09,+.39]$ & .025 & $+.01\,[-.13,+.15]^\ast$ \\
CosyVoice~2 & .38 & .89 & $+.51\,[+.36,+.67]$ & $<\!10^{-5}$ & $-.20\,[-.35,-.05]$ \\
Irodori     & .47 & .85 & $+.38\,[+.26,+.53]$ & $<\!10^{-4}$ & $-.19\,[-.30,-.07]$ \\
T5Gemma     & .60 & .93 & $+.33\,[+.19,+.49]$ & .002 & $-.26\,[-.41,-.09]$ \\
\midrule
Sarashina$^\ddagger$ & .67 & .81 & $+.15\,[-.02,+.31]$ & .17 & \\
\bottomrule
\end{tabular}
\end{table}

\textbf{Results.} Four admitted raters judged everything. Catch trials $8/8$, $8/8$, $7/8$ and $6/8$, duplicate consistency $23/23$ three times and $22/23$, none excluded; Fleiss' $\kappa$ over $441$ stimuli is $0.89$, $13$ tied stimuli excluded. All four tag-against-kana contrasts of Table~\ref{tab:subj} are significant in the family (discordant words $14$ vs.\ $3$, $23$ vs.\ $0$, $18$ vs.\ $0$ and $15$ vs.\ $1$). Against no edit on Sarashina, the tag is ahead but not significantly: unedited, this base already says two-thirds of these words right.

\subsection{Naturalness}
\label{sec:nat}
On words that needed no edit, the edit (tag) costs naturalness on three backbones and nothing on Sarashina (Table~\ref{tab:subj}). Following the critique of absolute rating scales~\cite{kirkland2023mospit}, listeners choose the more natural of the \emph{same sentence} with and without the edit, or no difference, on the easy set (\S\ref{sec:exp}; $50$ pairs per backbone, with one held-out voice each; both renditions come from a single session under one guard), where the edit leaves the perceived word unchanged. A pair with an inaudible clip (peak under $0.05$) is excluded before shipping: none were, so $200$ pairs ship in four groups of $50$, one per sitting. The easy-set tags were reviewed by the author before the edited ones were synthesized. Fixed in advance: a minimum of $15$ judgments per pair, limited to one session per account, and removal of any session whose answers come faster than its shortest pair can be played. The statistic is the share preferring the edited (tag) rendition minus the share preferring the no-edit one; ties are split evenly, with a pair- and rater-clustered bootstrap and a registered non-inferiority margin of $0.15$ at the lower $95\%$ limit.

\textbf{Results.} ($108$ sittings started, $102$ complete; $5{,}116$ judgments on $200$ pairs after two speed-rule exclusions, $24$ to $27$ per pair). The registered non-inferiority margin of $0.15$ is met for Sarashina alone (Table~\ref{tab:subj}, right). The pairs in which both renditions yield the same reading give the same picture ($+0.03$, $-0.22$, $-0.21$ and $-0.19$). CosyVoice~2 has a failure of its own, in every arm: it truncates the final mora of the sentence (absent from the transcripts for $29$ of $50$ kana and $45$ of $50$ tagged listening stimuli). The design is the least favorable one for an edit: every word in this case is one that the base already gets right, so any edit can only make things worse; by contrast, in use, only words the base gets wrong are recorded (\S\ref{sec:deploy}).

\section{Conclusion}
\label{sec:conclusion}
A pronunciation and accent control module can be added to a released TTS backbone using a small text inventory and self-generated supervision, without any recorded speech: distilling the model against itself on words it already says correctly yields joint reading-and-accent control for words it has never seen, on all four backbones. A registry applies it wherever a registered word appears. When judged by a selected native listener who judges reading and accent together, the tag says $0.80$ to $0.93$ for difficult words, whereas plain kana yields $0.38$ to $0.60$. On three backbones, listeners still prefer the unedited rendition of a word that needed no edit. The mechanism is the point: A model that already exhibits the behavior can be trained to elicit it using only its own outputs, as long as those outputs are first checked for any additional lessons they might impart. \textbf{Limitations.} One language; the rate rule followed an earlier round's failure; four raters.

\vfill\pagebreak

\section*{Acknowledgments}
\textbf{Funding and conflicts of interest.} No funding was received for this study; compute (the author's own hardware and cloud GPUs rented through the author's sole proprietorship) and the listening studies were paid for by the author. The author is affiliated with KOWRO Co., Ltd., a speech-synthesis company that provided no funding, data, or compute for this work. The method evaluated is the author's own; every accent judgment is therefore made blind, by screened raters, under a pre-registered design.

\textbf{Use of generative AI.} Generative AI (Claude, Anthropic) was used under the author's direction and correction to refine sentences in this paper, to draft and edit the pre-registration, to write the experiment, evaluation, and analysis code and prepare its release, and to produce Fig.~\ref{fig:both}. The author approved every design and analysis decision recorded in the registration, checked the results, and is responsible for all content.

\section*{Compliance with ethical standards}
This study was conducted in accordance with the principles of the Declaration of Helsinki; the author has no institutional ethics committee to which to apply, so no committee review was sought. The listening studies collect only non-invasive judgments on synthesized speech. No audio, biometric, or personally identifiable data are gathered. Participants receive a written description of the task, data use, and compensation, and give informed consent; participation is voluntary, may be discontinued at any time, and a discontinued session is still paid. Every candidate takes the same screening and is paid for it, whatever the outcome.

\bibliographystyle{IEEEbib}
\bibliography{refs}

\end{document}